\documentclass{iaus}
\usepackage{graphicx}
\usepackage{natbib}

\title[Galactic dynamo simulations]{Galactic dynamo simulations}

\author[D.~Elstner, O.~Gressel \& G.~R\"udiger]%
{Detlef Elstner, Oliver Gressel and G\"unther R\"udiger}

\affiliation{Astrophysikalisches Institut Potsdam, An der Sternwarte
16, D-14482 Potsdam, Germany
  \break email: elstner@aip.de}

\pubyear{2009}
\volume{259}
\pagerange{100--100}
\date{Dec. 10, 2008 and in revised form ??}
\jname{Cosmic Magnetic Fields: From Planets,to Stars and Galaxies}
\editors{K.G. Strassmeier, A.G. Kosovichev \& J.E. Beckman, eds.}

\newcommand{\Myr}{\,\rm Myr}
\newcommand{\Gyr}{\,\rm Gyr}
\newcommand{\pc}{\,\rm pc}
\newcommand{\kpc}{\,\rm kpc}
\newcommand{\kms}{\,\rm km\,s^{-1}}
\newcommand{\muG}{\,\rm \mu G}
\newcommand{\Gauss}{\,\rm G}
\newcommand{\Beq}{B_{\rm eq}}
\newcommand{\AuA}{A\&A}
\newcommand{\MN}{MNRAS}
\newcommand{\AN}{AN}
\newcommand{\ApJ}{ApJ}
\newcommand{\ARAA}{ARAA}

\begin{document} 

\maketitle

\begin{abstract}
Recent simulations of supernova-driven turbulence within the ISM
support the existence of a large-scale dynamo. With a growth time of
about two hundred million years, the dynamo is quite fast -- in
contradiction to many assertions in the literature. We here present
details on the scaling of the dynamo effect within the simulations
and discuss global mean-field models based on the adopted turbulence
coefficients. The results are compared to global simulations of the
magneto-rotational instability. 
\end{abstract}

\firstsection 

\section{Introduction}

Large-scale coherent magnetic fields are observed in many spiral
galaxies \citep{Beck96,Beck08}. It is now widely accepted that a
turbulent dynamo process is responsible for the sustained
amplification of the mean magnetic field within galaxies. The
turbulence driven by supernova explosions together with the
differential rotation should in fact lead to the action of the
classical $\alpha\Omega$~dynamo \citep{Parker71}. First steps to
globally model flat objects in an ellipsoidal geometry were done by
\citet{Stix75,White}. Starting from the early nineties, first
multidimensional numerical mean-field models assuming a disk geometry
appeared \citep{Donner,Elstner92}. These models applied simple
estimates for the turbulent transport coefficients in the mean-field
induction equation.

\citet{RuedigerKit93}, in the framework of second order correlation
approximation (SOCA), derived the full $\alpha$~tensor and turbulent
diffusivity for stratified turbulence. \cite{FroehlichSchultz}, under
the assumption of hydrostatic equilibrium, computed the turbulent
velocity distribution $u'(z)$ based on the observed density
stratification and an observationally motivated gravitational
potential from stars and dark matter. Applying SOCA theory, one could
now specify the full $\alpha$~tensor, resulting in a model with the
correlation time $\tau_{\rm c}$ of the turbulence as the only free
parameter.

With the increased computer power of the nineties, \citet{Kais93} and
\citet{Ziegler96}, in an alternative approach, computed the turbulence
coefficients by averaging the results from simulations of a single
supernova remnant. It, however, turned out that the values from
isolated explosions were about one magnitude smaller then the
estimated numbers from the previous considerations. Similar results,
based on semi-analytical models, were obtained by \citet{Ferriere98}.
Extending the investigation to super-bubbles and including the
stratification of the ISM, she finally found turbulence coefficients
of expected magnitude. However, with a huge escape velocity, which
could switch off the dynamo process. \citet{FerriereSchmitt} used
these results as input for an axisymmetric dynamo model of our
Galaxy. They reached growth times of about $450\Myr$ but with a ten
times larger toroidal field than the radial component, i.e., with a
rather low pitch angle.

A lot of models, but mostly axisymmetric, were published during that
period. Most of the models were investigating different aspects of the
turbulence, looking also for parameters where non-axisymmetric
solutions could appear. Another application was the effect of radial
flows on the dynamo properties driven by the Maxwell stress of the
dynamo generated magnetic field. Models including a wind were
investigated, especially to explain vertical fields above the disk --
notably \citet{Brandenburg93,Elstner95}. Both models used different
assumptions about the halo diffusivity and the wind structure. In the
model of \citet{Brandenburg93}, diamagnetism was included, and because
of the positive turbulence gradient, the pumping was not an escape
velocity but an inward transport term. In \citet{Elstner95}, a passive
halo with low conductivity was assumed. These models already showed a
fast growth time below $0.1\Gyr$. All these models could in principle
explain the main features of the observed magnetic fields in nearby
spiral galaxies. But the connection between optical and magnetic
spirals was still unclear.

With the advent of the new century, full-blown 3D models became
feasible, and one could now attack the spiral structure. Simple
artificial velocity models adopted from density-wave theory were now
taken into account. One spectacular observational result was the
anti-correlation of optical and magnetic arms, found by
\citet{BeckHoernes}. A turbulent dynamo could reproduce such a
behaviour due to the different properties of the turbulence in the
spiral arms and in the inter-arm regions -- see \citet{Rohde98},
\citet{Shukurov98} and \citet{RohdeBeck99}.

More elaborated models used the velocity from numerical galaxy models
computed with SPH or sticky-particle methods \citep{Elstner00}. Models
for spiral galaxies produced no inter-arm fields by the large-scale
spiral motion alone. Only with a variation of the turbulence, one
could have inter-arm fields. The situation was different for bar
galaxies \citep{Otmianowska}. In these simulations, one could observe
inter-arm magnetic fields in the outer spiral arm, and the field
lagged behind the optical arm \citep[cf.]{Moss98}. More recently,
\citet{Beck05} and \citet{Moss07} conducted models for barred galaxies
with stationary flows.

Another problem was raised in the context of the so-called
catastrophic quenching: Because of helicity conservation, in the ideal
MHD case, the $\alpha$~term will be quenched (with the square-root of
the magnetic Reynolds number) if there is no additional transport of
magnetic helicity in the box. Under the assumption of a galactic wind
or fountain flow, these problems may disappear. But a too strong wind
would also result in a suppression of the dynamo \citep{Sur07}. Recent
investigations have shown that already the galactic shear flow can
suppress the catastrophic quenching of the dynamo
\citep{Brandenburg08,kaepylae08}.

The recent box model of the turbulent ISM driven by multiple clustered
supernova explosions \citep{Gressel0}, for the first time demonstrated
the presence of a dynamo process by direct numerical simulations. It
turned out that the turbulence coefficients derived from the SOCA
models are in good agreement with the new results from the direct
simulations. In a similar way, the box simulations of \cite{Hanasz04}
have shown an exponential growth of the magnetic field. Their
simulations, however, do not resolve the supernova remnant in itself.
Instead, the explosion site only serves as a source for the released
cosmic ray energy. Due to the fast diffusion of cosmic rays, the
pressure redistribution is fast enough to allow for similarly short
growth times of several $100\Myr$. The role of a magnetic instability
of the Parker type could be essential for this process, too.

In general, the effect of magnetic instabilities in releasing kinetic
energy should be considered as a possible mechanism for magnetic field
amplification in galaxies -- especially in regions with low star
formation activity. The primary candidate of this type is the
magneto-rotational instability (MRI), which has a growth time on the
order of the galactic rotation frequency and thus can become efficient
on very short timescales. With todays computational resources, global
models of the MRI are already within reach
\citep{Dziourkevitch04,Nishikori06}. However, these models still have
to assume quite strong initial fields to properly resolve the fastest
growing modes of the instability.

\section{The pitch angle problem}\label{sec:pitch}

When it comes to the explanation of the observed magnetic fields in
spiral galaxies, there is one main reason for favouring the dynamo
hypothesis over the primordial field hypothesis, namely the existence
of fields -- of equipartition strength -- with a large pitch angle.
If the field amplification were mainly due to the shearing motion of
the differential rotation, the field would have to be wound up several
times to explain its observed strength. This, however, would lead to
an amplification of the toroidal field only, resulting in a
configuration with an intrinsically small pitch angle.

For the dynamo scenario, which furthermore implies exponential growth,
we additionally have an amplification of the poloidal field. The pitch
angle $p$ of the growing field in the kinematic regime of the
$\alpha\Omega$~dynamo can be estimated from the dimensionless dynamo
numbers $C_\alpha$ and $C_\Omega$ (for a definition see below) via
\begin{equation}
  p=\arctan\left(\sqrt{\frac{C_\alpha}{C_\Omega}} \right)\,.
\end{equation}
This leads to values between $5^\circ$ and $30^\circ$, for reasonable
values of $C_\alpha=5$--$10$ and $C_\Omega=25$--$800$. For a given
rotational velocity, the angular frequency is defined by the turnover
radius, i.e., the point where the rotation curve becomes flat. Small
turnover radii lead to rather large $C_\Omega$ and thus to low pitch
angles. In the saturated regime, the situation becomes worse -- at
least for the conventional $\alpha\Omega$~dynamo. This is because
$\alpha$ will then be reduced by non-linear effects and the pitch
angle fades.

Direct simulations of the ISM turbulence show a possible way out of
this dilemma, namely by another saturation process, which we will
demonstrate in the following section. Moreover, the action of the
magneto-rotational instability may be a reasonable candidate for the
explanation of large pitch angles as they are observed in maps of
polarised synchrotron emission for instance in the ringed galaxy
NGC~4736 \citep{Chyzy08}.

\section{Global models with turbulence coefficients from direct simulations}
\label{sec:meanbox}

Global simulations of interstellar turbulence remain out of reach, at
least if one is interested in resolving the relevant
scales. Therefore, we apply a hybrid approach, where we consider the
mean induction equation
\begin{equation}
  {\partial \vec{B} \over \partial t}
  = {\rm curl}(\vec{u} \times \vec{B} + \alpha\circ \vec{B}
    -{\eta}_{\rm T} \circ {\rm {curl}}{\vec{B}})
\end{equation}
for the mean magnetic field $\vec{B}$ and apply closure parameters
obtained from direct simulations. The background rotation serves as
the main part of the mean flow
\begin{equation}
  u_\varphi=
  r \Omega_0 \left(1 + \left({r \over r_\Omega}\right)^2 \right)^{-1/2}\,,
  \label{eq:Brandt}
\end{equation}
and is supplemented by a vertical wind $u_z= u_0 z$ which is a
result of the density-stratified turbulence driven by SNe (cf.
Fig.~2 in Gressel et al. 2009). The test-field method, furthermore,
yields an $\alpha$~tensor of the form
\begin{equation}
  \alpha = \alpha_0 \left( \matrix{\alpha_{rr} & -\gamma_z & 0 \cr
    \gamma_z & \alpha_{\varphi\varphi} & 0  \cr
    0 &  0 & \alpha_{zz} \cr } \right)\,,
\end{equation}
where the dependence of these quantities on the height $z$ is modelled
according to the profiles in \citet{Gressel0}.

For a realistic global galaxy model one further needs the radial
dependence of the relevant effects. The radial profiles of the
rotational frequency, SN~rate, gravitational potential and density
are in principle known. By means of a parameter study, we try to
obtain the $\alpha$~effect from our box simulations, accordingly.
From these, we find a dependence of the diagonal terms of the
$\alpha$~tensor on rotation, namely $\alpha_{\varphi\varphi} \propto
\Omega^{0.5}$. This is weaker than expected from SOCA theory, which
predicts a scaling linear with $\Omega$ and only a minor rotational
quenching. Consistent with SOCA, we find the turbulent diamagnetism
$\gamma_z$ to be nearly independent of $\Omega$. For the turbulent
diffusivity $\eta_{\rm t}$, we apply only horizontal components
which also vary according to $\Omega^{0.5}$. The dependence on the
supernova rate is given in Gressel et al. (2009).

After proper normalisation with a length of ${l}=1\kpc$ and a
diffusion time $\ t_{\rm D}={l}^2/\eta_0$ with $\eta_0 = 1
\kpc^2\Gyr^{-1}$, the dynamo is characterised by the Reynolds
numbers
\begin{equation}
  C_\alpha=\alpha_0 {l}/\eta_0 \,, \quad
  C_\Omega=\Omega_0 {l}^2/\eta_0 \quad \textrm{and} \quad
  C_w=u_0 {l}/\eta_0\,.
\end{equation}

\begin{figure}
    \includegraphics[width=0.95\columnwidth]{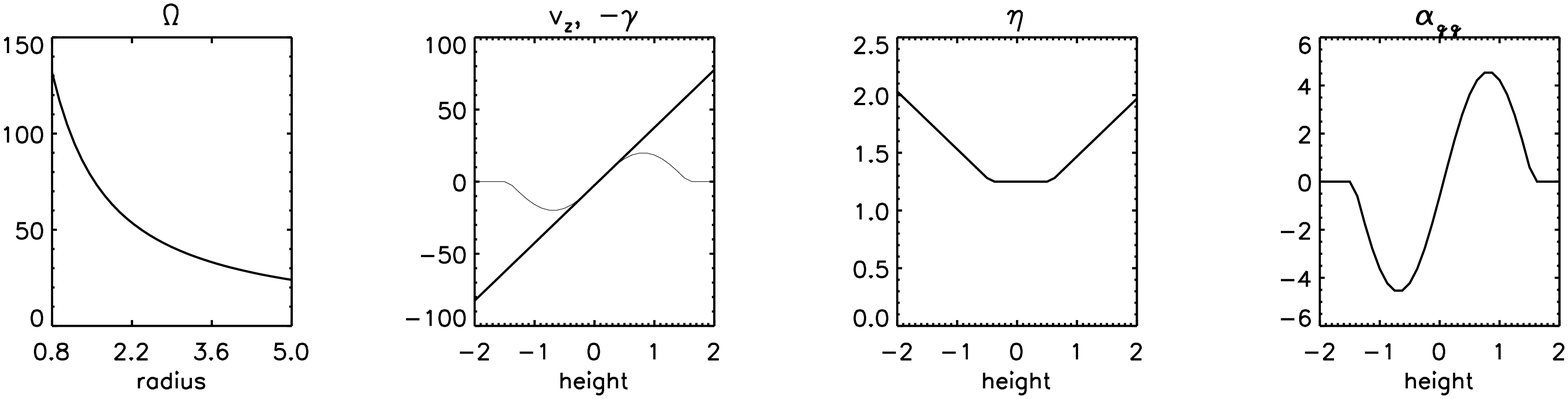}
    \caption{Radial dependence of $\Omega$
             and vertical profiles of the fountain flow and turbulent
      transport coefficients.}
  \label{fig:model}
\end{figure}
We assume a disk of radius ${R}=5\kpc$ and with a height of
${H}=2\kpc$, according to our box dimension in the direct simulations.
We approximate the profiles of the turbulence coefficients by smooth
functions (see Fig.~\ref{fig:model}). The vertical dependence of the
$\alpha$~tensor can simply be reproduced by a sine function
\begin{equation}
  \alpha(z) \propto \sin(z \upi / {\rm h}_\alpha )
\end{equation}
with a scale height ${h}_\alpha=1.5 \kpc$. The mean vertical velocity
is linearly growing with height. Only the diagonal terms of the
$\alpha$~tensor have a radial dependence proportional to $\Omega$,
whereas the other terms are constant. We, furthermore, use a simple
$\alpha$~quenching
\begin{equation}
  \alpha=\alpha_0 / (1 + B^2/\Beq^2)
\end{equation}
for the full tensor, i.e., including the diamagnetic term $\gamma_z$.

In our first run, we neglect turbulent diamagnetism and mean vertical
velocity, as it is done in the classical $\alpha\Omega$~dynamo. In
this model, we find a growth time of $700\Myr$, a field strength for
the toroidal component $B_{\varphi}= 20 \Beq$ and the radial component
$B_{r}= 0.6 \Beq$. If we neglect only the fountain flow, the field
decays. For the \textit{real} model, we find a growth rate of
$250\Myr$ -- similar to the box simulation. The field now saturates
earlier with $B_{\varphi}= 1.1 \Beq$ and $B_{r}= 0.3 \Beq$.

\begin{figure}
  \resizebox{4.4cm}{3.5cm}{\includegraphics{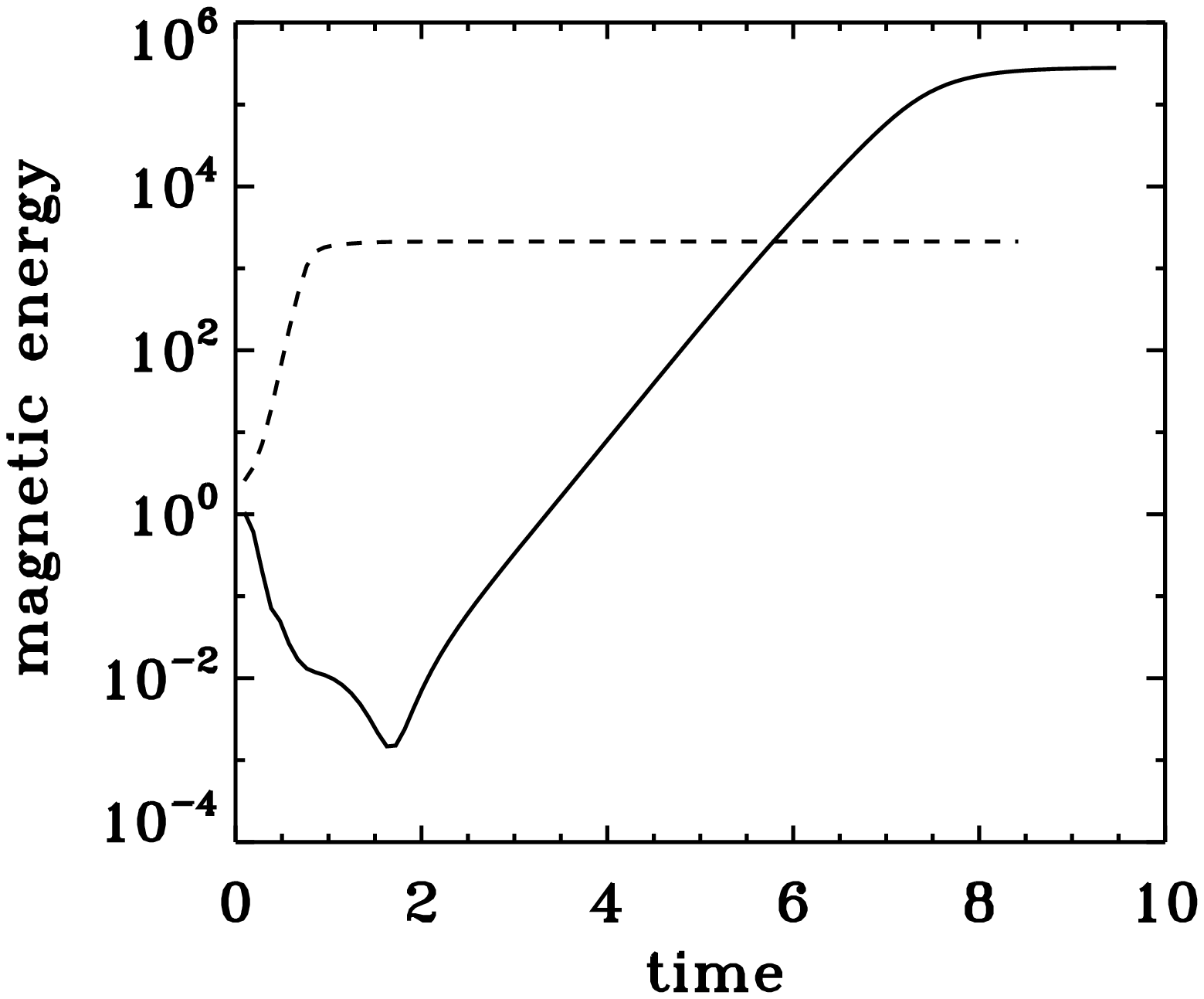}}
  \resizebox{4.4cm}{3.5cm}{\includegraphics{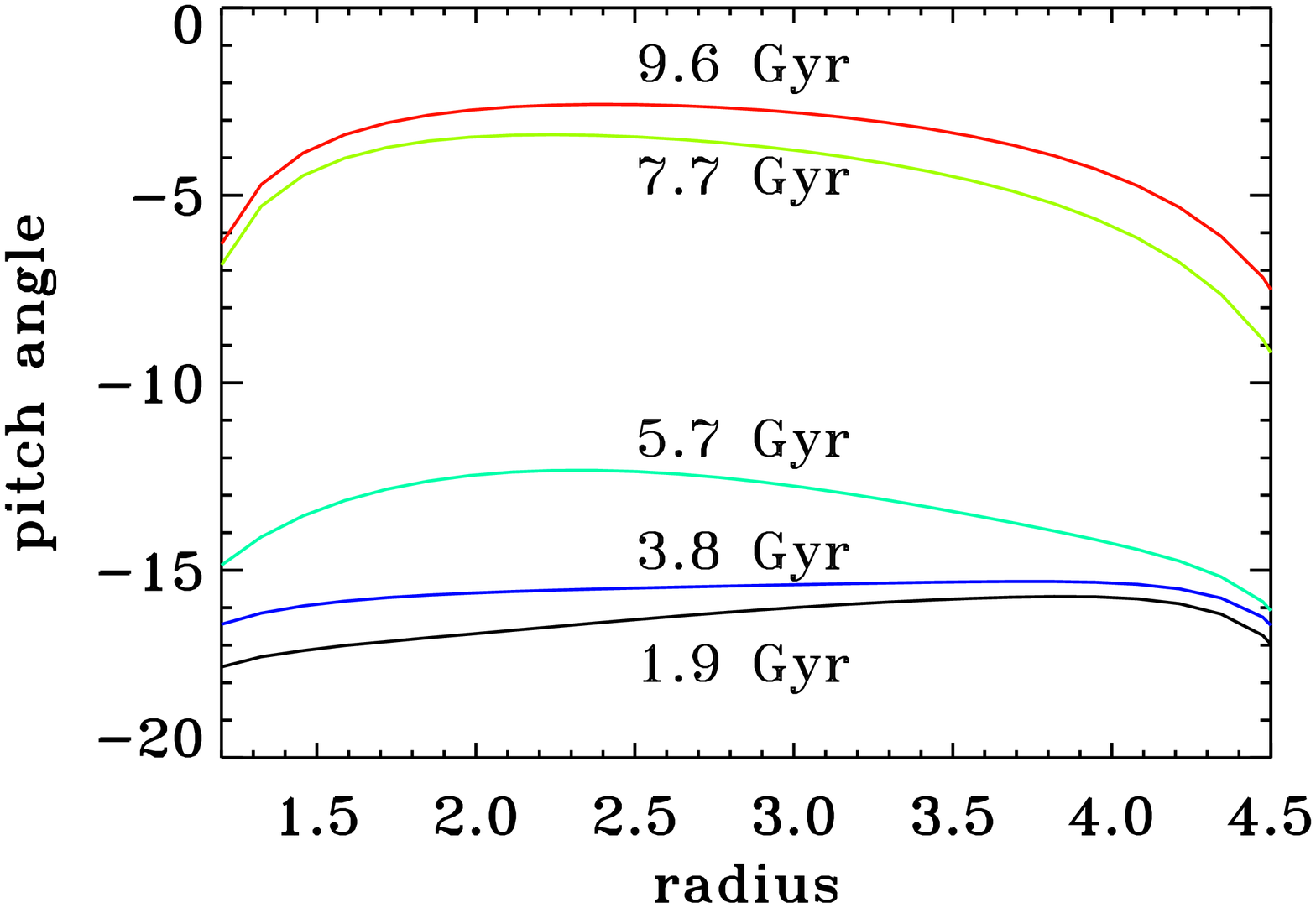}}
  \resizebox{4.4cm}{3.5cm}{\includegraphics{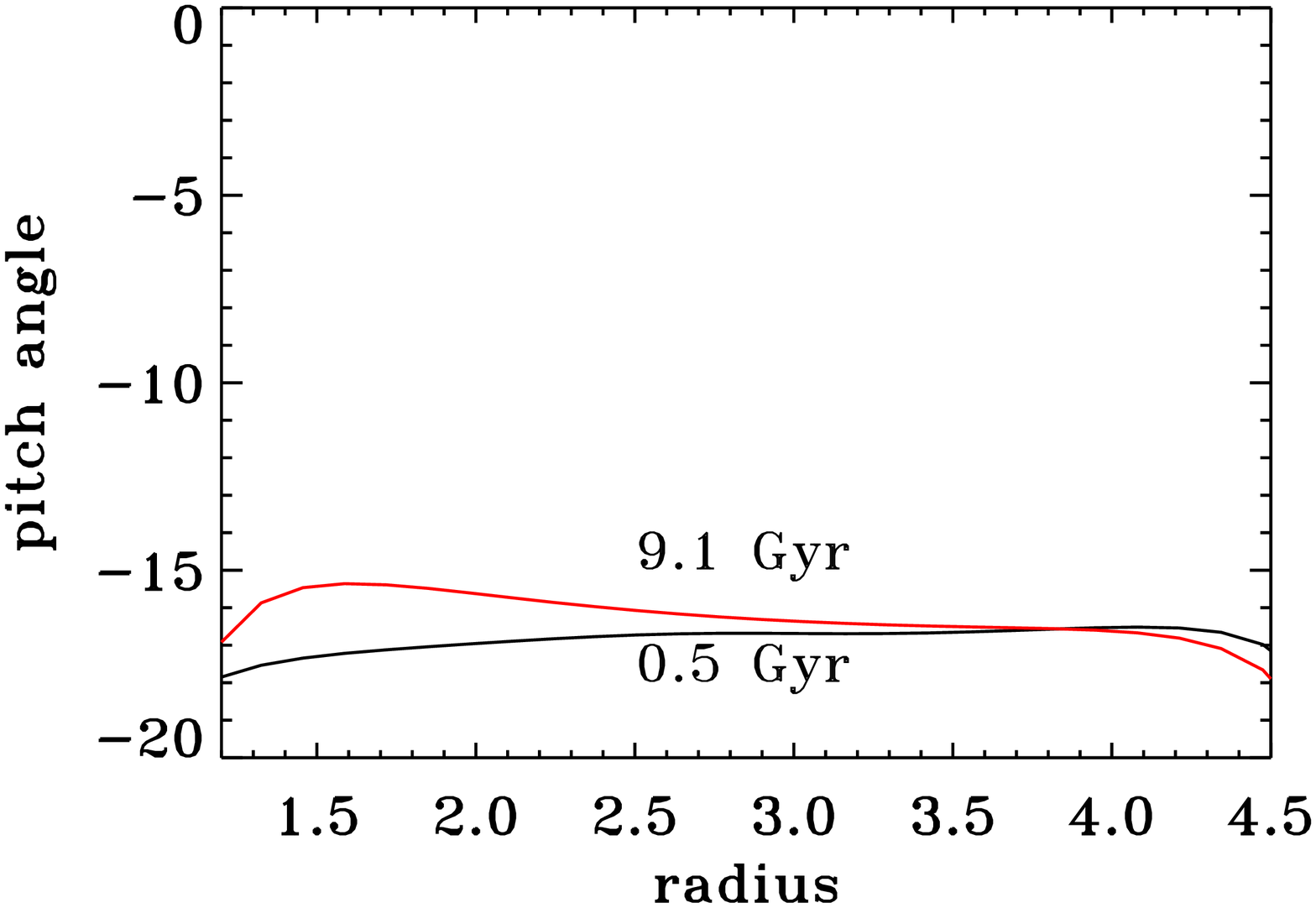}}
  \caption{Left: Time evolution of magnetic energy for model A, with
                 wind (dashed) and model F, without wind and pumping
                 (solid). Middle: Time evolution from $1.9\Gyr$ until
                 $9.6\Gyr$ of the radial pitch angle distribution for
                 model F. Right: Pitch angle for model A at $0.5\Gyr$
                 and $9.1\Gyr$.}
  \label{fig:pitch}
\end{figure}

The final state is not determined by the usual $\alpha$~quenching in
this case. Instead, a combination of fountain flow and quenching of
the diamagnetic term leads to the saturation of the dynamo. This
means, the dynamo stops growing because of field losses by the
fountain flow. As a nice consequence of this non-linear feedback, the
pitch angle remains large in the saturated regime
(see Fig.~\ref{fig:pitch}).

\begin{figure}
  \begin{center}
    \resizebox{6.5cm}{6.5cm}{\includegraphics{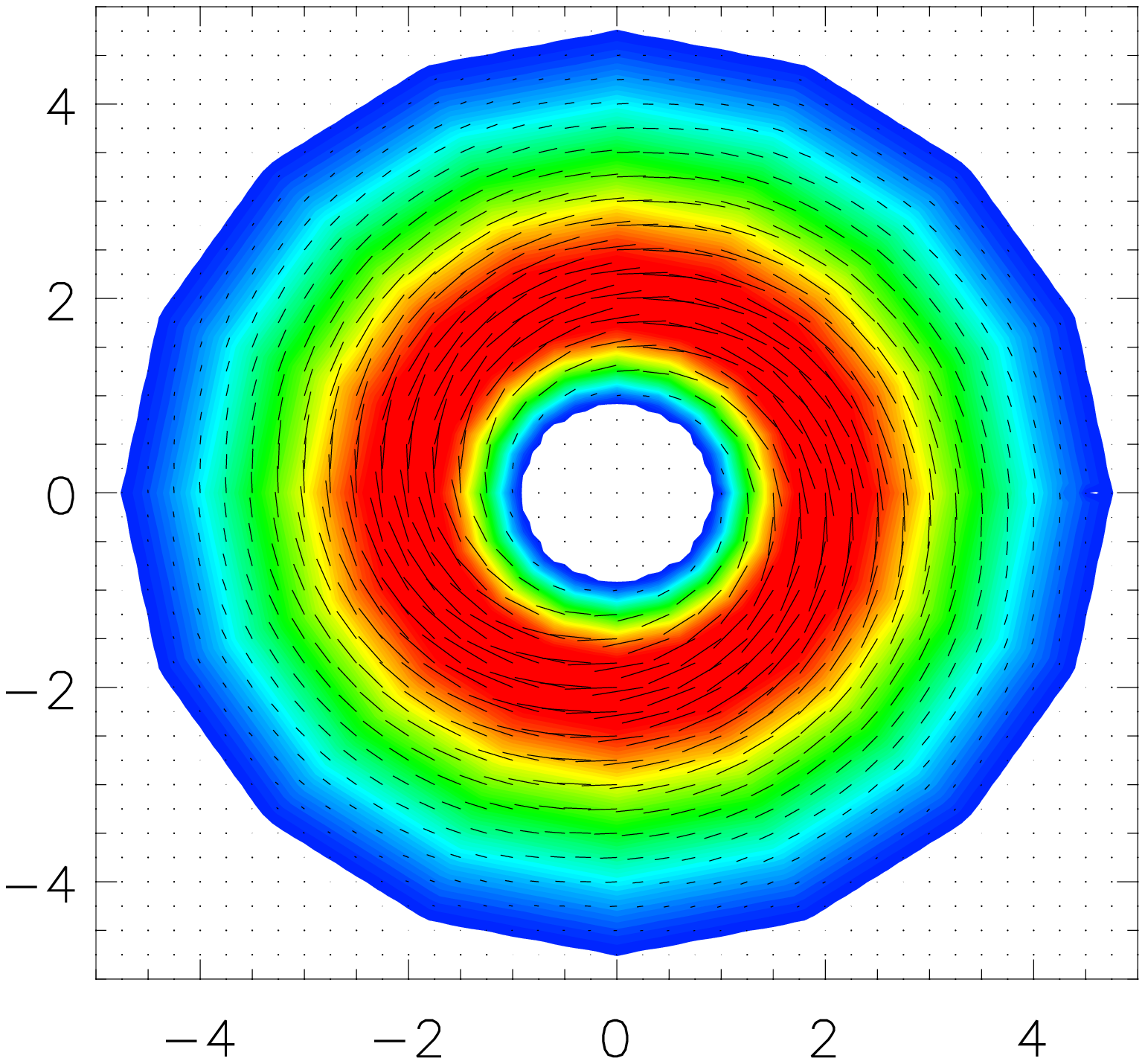}}
    \resizebox{6.5cm}{6.5cm}{\includegraphics{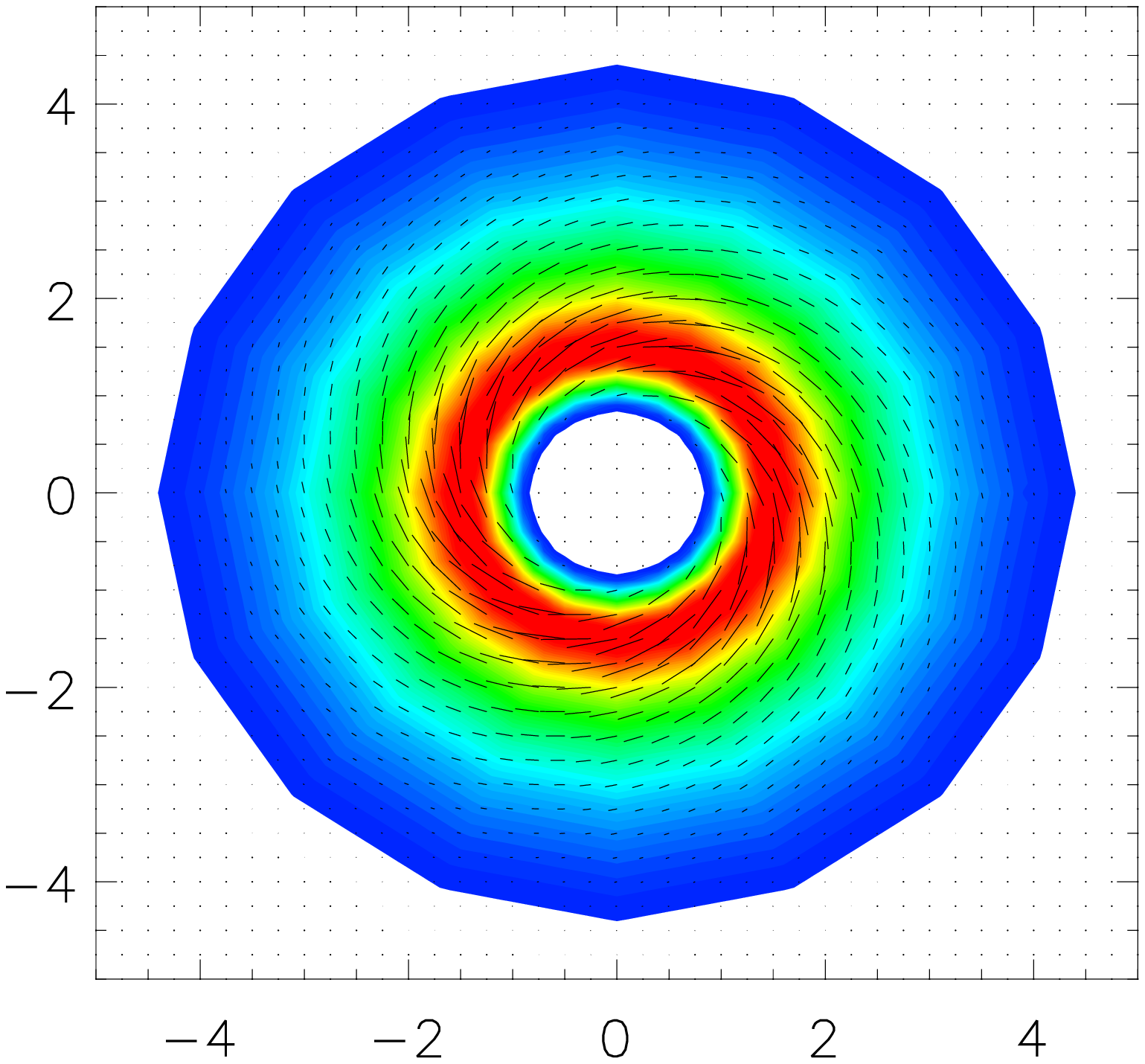}}
  \end{center}
  \caption{Magnetic vectors as observed in polarised emission
  (neglecting Faraday rotation) for the final fields in model F (left)
  and model A (right)} 
  \label{fig:polmap}
\end{figure}

Moreover, in Fig.~\ref{fig:polmap} we present directions of the
magnetic field resulting from polarised emission of synchrotron
radiation, where we assumed a scale height of $500\pc$ for the
relativistic electron distribution.

In Table~\ref{tab:pitch}, the pitch angle is given for a choice of
different models. Comparing $p=\arctan(\sqrt{C_\alpha / C_\Omega})$ with the
initial pitch angle $p_0$ during the kinematic growth phase, we find
reasonable agreement. The final value $p_\mathrm{final}$ in the
saturated state for the first three models is similar to the value
during the kinematic growth phase.

In order to have a higher pitch angle for our rotation law with
$r_\Omega=1\kpc$, we need larger values of $C_\alpha$ and,
consequently, a larger vertical outflow. Otherwise the
$\alpha$~quenching would reduce the final pitch angle -- as is the
case in model E. This increase of the mean outflow velocity with
increasing SN~rate is indeed observed in the box simulations (cf.
Gressel et al. 2009).  For model F, one can see a dramatic reduction
of the pitch angle due to the missing mean vertical velocity and
diamagnetic pumping.
 \begin{table}\def~{\hphantom{0}}
 \begin{center}
  \caption{Pitch angles for different dynamo models}
  \label{tab:pitch}
  \begin{tabular}{lrrrrrrrc}\hline
      & $C_\alpha$  & $C_\Omega$   &  $p$ &  $p_0$ &
      $p_\mathrm{final}$ & $v_\mathrm{rot}$ & $u_0$ &
      $\gamma/\alpha_{\varphi\varphi}$  \\
      \hline
       A & 5  & 100 & -13 & -16 & -16 & 100 & 20 & 2\\
       B & 5  & 200 & -9  & -10 & -10 & 200 & 20 & 2\\
       C & 5  &  50 & -18 & -17 & -15 &  50 & 20 & 2\\
       D & 25 & 100 & -27 & -28 & -18 & 100 & 20 & 2\\
       E & 25 & 100 & -27 & -28 & -27 & 100 & 40 & 2\\
       F & 5  & 100 & -13 & -16 &  -3 & 100 &  0 & 0\\
       \hline
  \end{tabular}
 \end{center}
\end{table}

\section{Magneto-Rotational Instability}\label{sec:mri}

Galaxies are known to be unstable with respect to the
magneto-rotational instability. This instability, discovered by
\citet{Velikhov59} and \citet{Chandrasekhar60}, has meanwhile been
studied in great detail in the astrophysical context \citep[see
e.g.][]{BalbusHawley91,KitchatinovMazur97}. \citet{SellwoodBalbus99} pointed
to the possibility of turbulence generation by the MRI in regions of
low star formation activity. \citet{Piontek} recently simulated the
MRI in a vertically stratified two-phase ISM box model.

\cite{Kitrue04} investigated the stability of global galactic
disks. For a disk threaded by a weak vertical field, they estimated a
field strength of $10^{-25}\Gauss$ for the onset of the
instability. Accordingly, the MRI should already be present during the
galaxy formation process -- and therefore a good candidate for the
seed field of a turbulent dynamo. The most unstable mode in their
idealised galactic disk model is of quadrupolar type. A very
interesting result is the large pitch angle of $45^\circ$ for the
fastest growing mode. For weak fields, the unstable modes have a small
spatial scale but with the property of a uniform large pitch
angle. This leads to a smooth map of magnetic vectors for the
polarised radio emission, but the Faraday rotation measure will be
small for these types of magnetic fields.

The instability, moreover, has a very short growth time proportional
to the rotation time. A proof for a large scale dynamo driven by the
MRI in the sense, that a small initial field can be amplified over
several magnitudes has not been given yet. This is because it is
numerical difficult to resolve the small spatial scales for the
unstable modes connected to the weak initial field. Nevertheless, there
are hints that the process is possible. Especially the turbulence
excited by MRI in a stratified disk gives a non-vanishing
$\alpha$~effect, but with negative sign in the northern hemisphere
\citep{Arlt, BrandenburgSokoloff02}.

Since we are interested in configurations with large pitch angles, we
here consider global models of a density-stratified galactic disk with
a uniform vertical field. Tests with toroidal fields did not produce a
comparable radial field component. For simplicity, we assumed an
isothermal disk where the rotation and vertical density stratification
is supported by the gravitational potential. Without magnetic fields
the configuration was tested to be in a stationary equilibrium over
$10\Gyr$ with only small velocity fluctuations below $0.01\kms$, which
are slowly decaying with time.

The applied density profile is
\begin{equation}
  \rho=\rho_0 \exp \left({-4z^2 \over z^2 + 3 h_\rho^2}\right)
\end{equation}
with the scale height $h_\rho=0.3\kpc$ and $\rho_0=10^{-24} {\rm g
cm}^{-3}$.  Note that the density minimum is about 0.02 $\rho_0$. The
rotation curve is, furthermore, defined by (\ref{eq:Brandt})
with the turnover radius $r_\Omega=0.35\kpc$ and $\Omega_0=
200\Gyr^{-1}$. The inner radius of our disk model is $r_{\rm
in}=1\kpc$ and the outer radius $r_{\rm out}=4\kpc$. The vertical
extent is $\pm5\kpc$.  Within $\pm1\kpc$, we use a constant grid
spacing with 256 grid points and a non-uniform grid for the outer
regions with 128 points at each side. This means, the inner disk has a
resolution of about $7\pc$ and for the outer disk we yield about
$80\pc$. This is consistent with a density variation of a factor 100,
in order to get a comparable growth of the unstable modes for a
uniform magnetic field strength within our restricted resolution.

The radial boundary is reflecting and the vertical boundary is reflecting for
the velocity only, while the tangential component of the magnetic
field and the gradient of the vertical component are set to
zero. Under these conditions, the vertical flux is conserved but a
development of horizontal flux is allowed.
\begin{figure}[h]
\vbox{
  \resizebox{6.8cm}{5cm}{\includegraphics{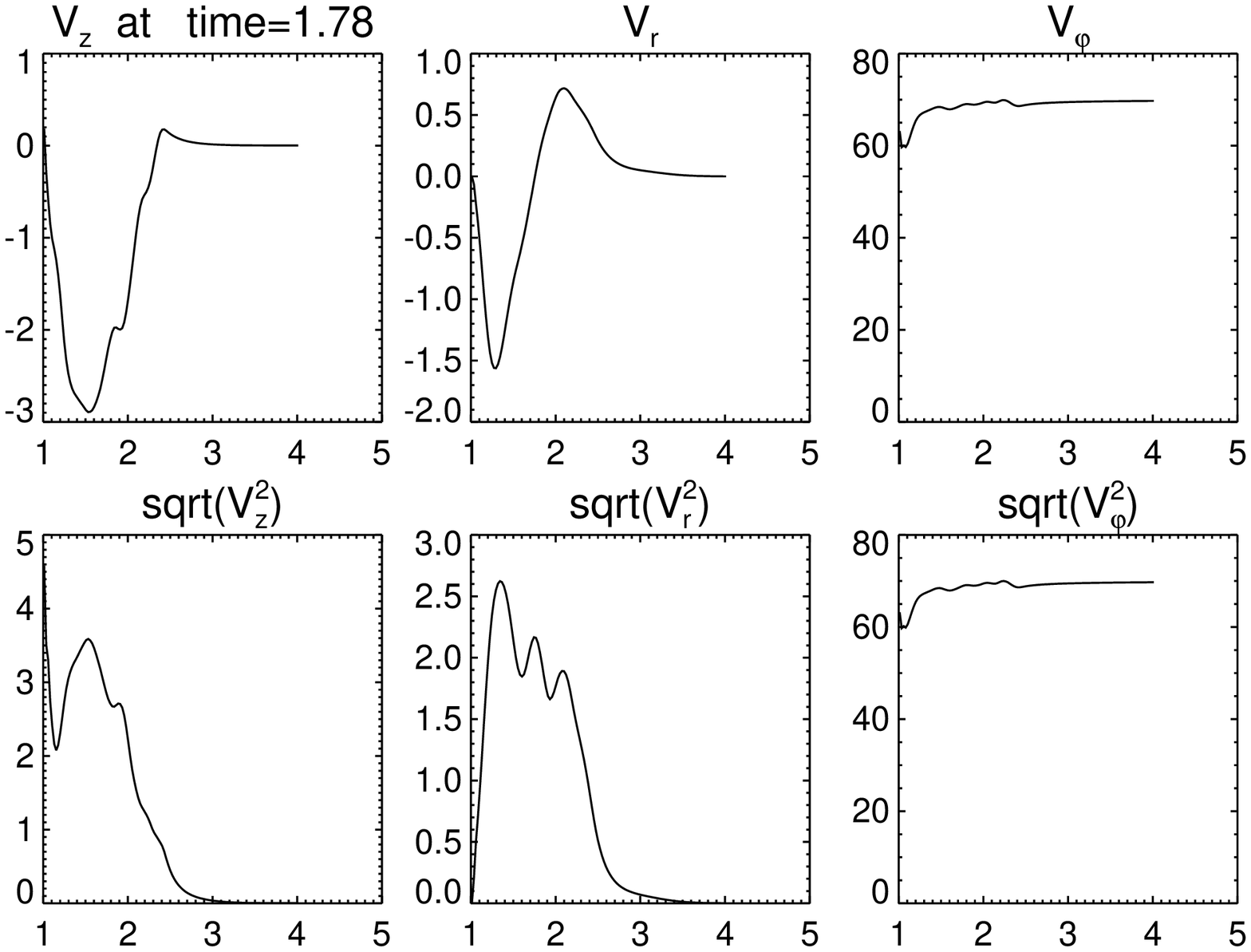}} \hfill
  \resizebox{6.8cm}{5cm}{\includegraphics{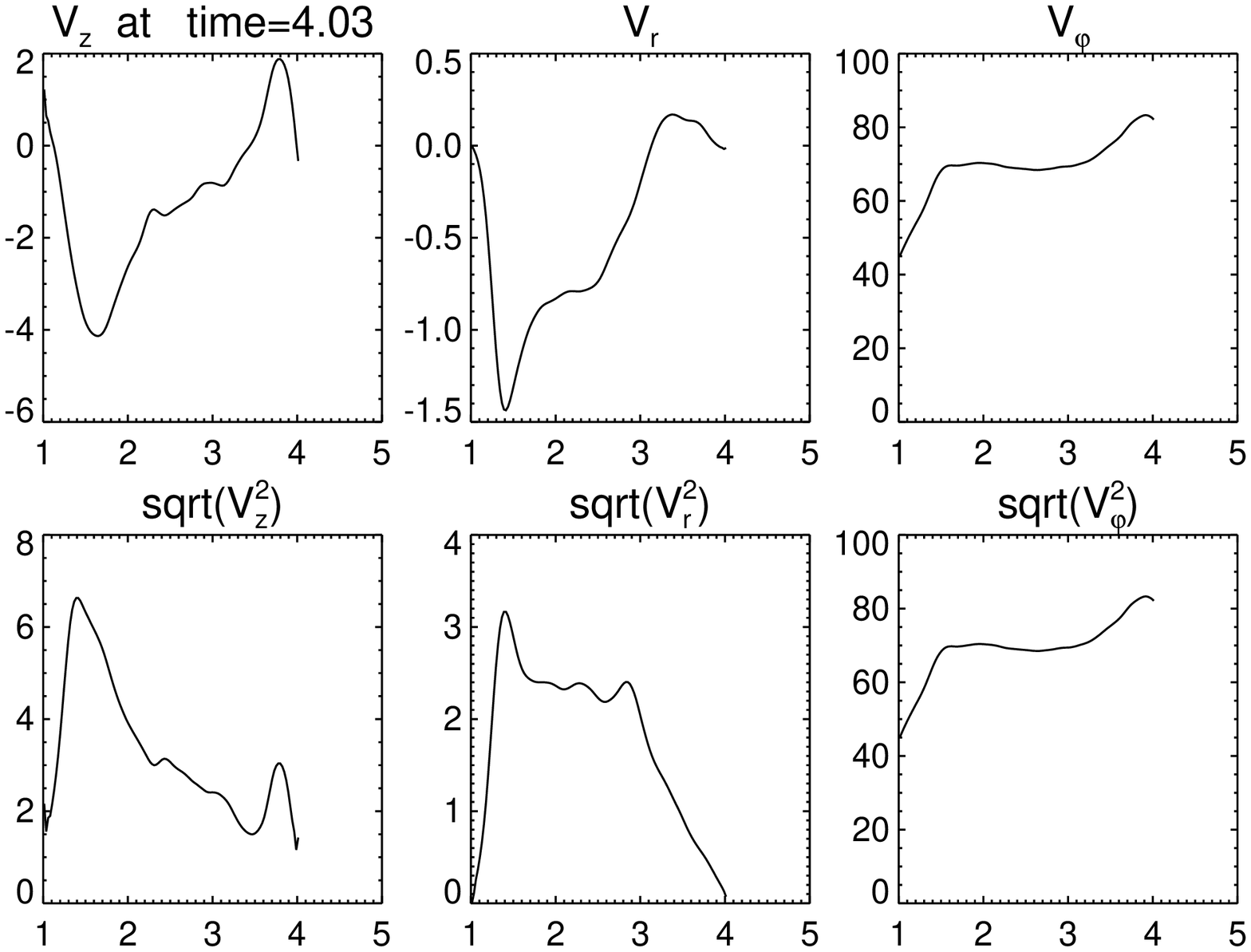}}
    }
   \vspace{0.3cm}
  \resizebox{6.8cm}{5cm}{\includegraphics{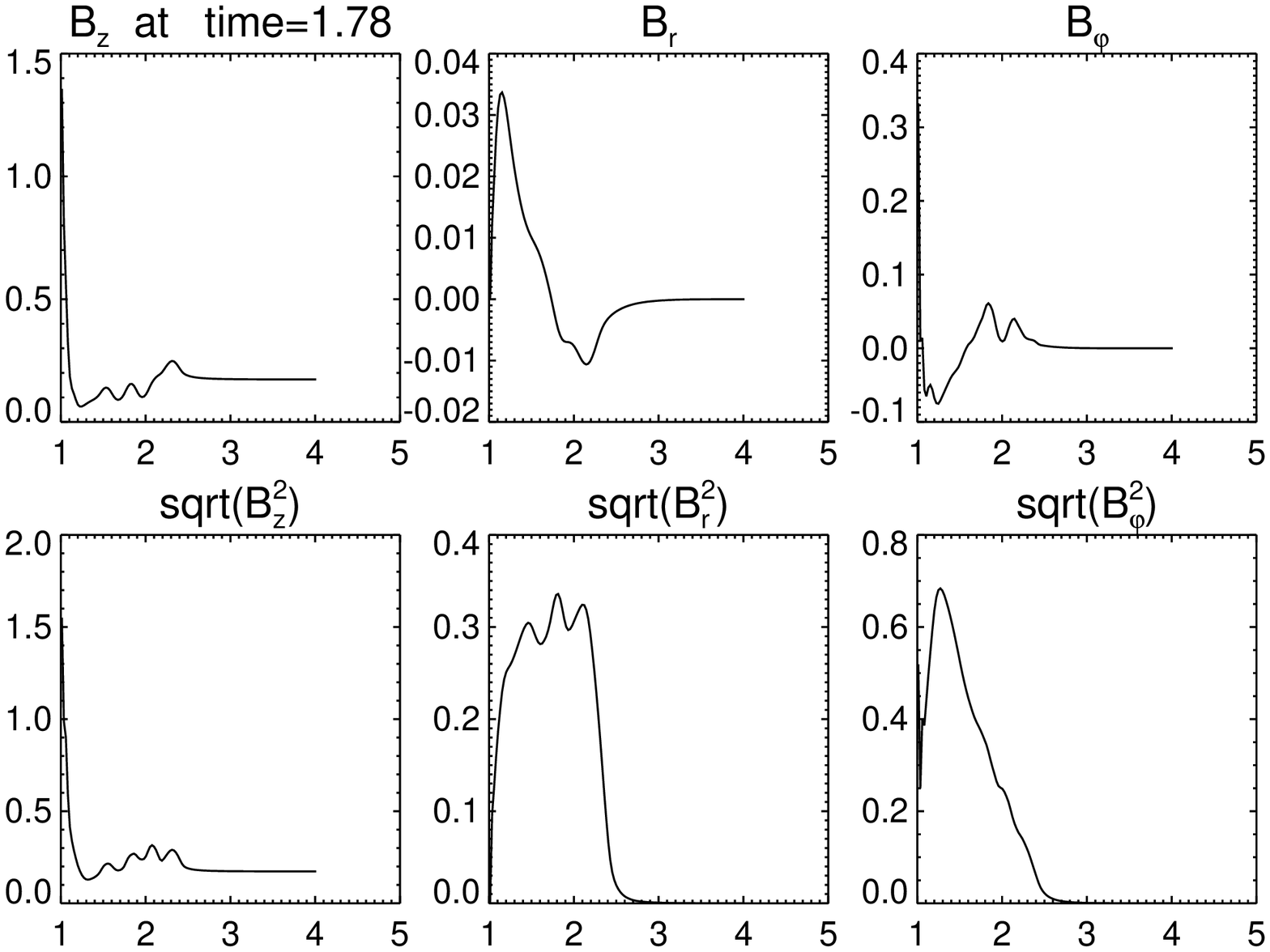}} \hfill
  \resizebox{6.9cm}{5cm}{\includegraphics{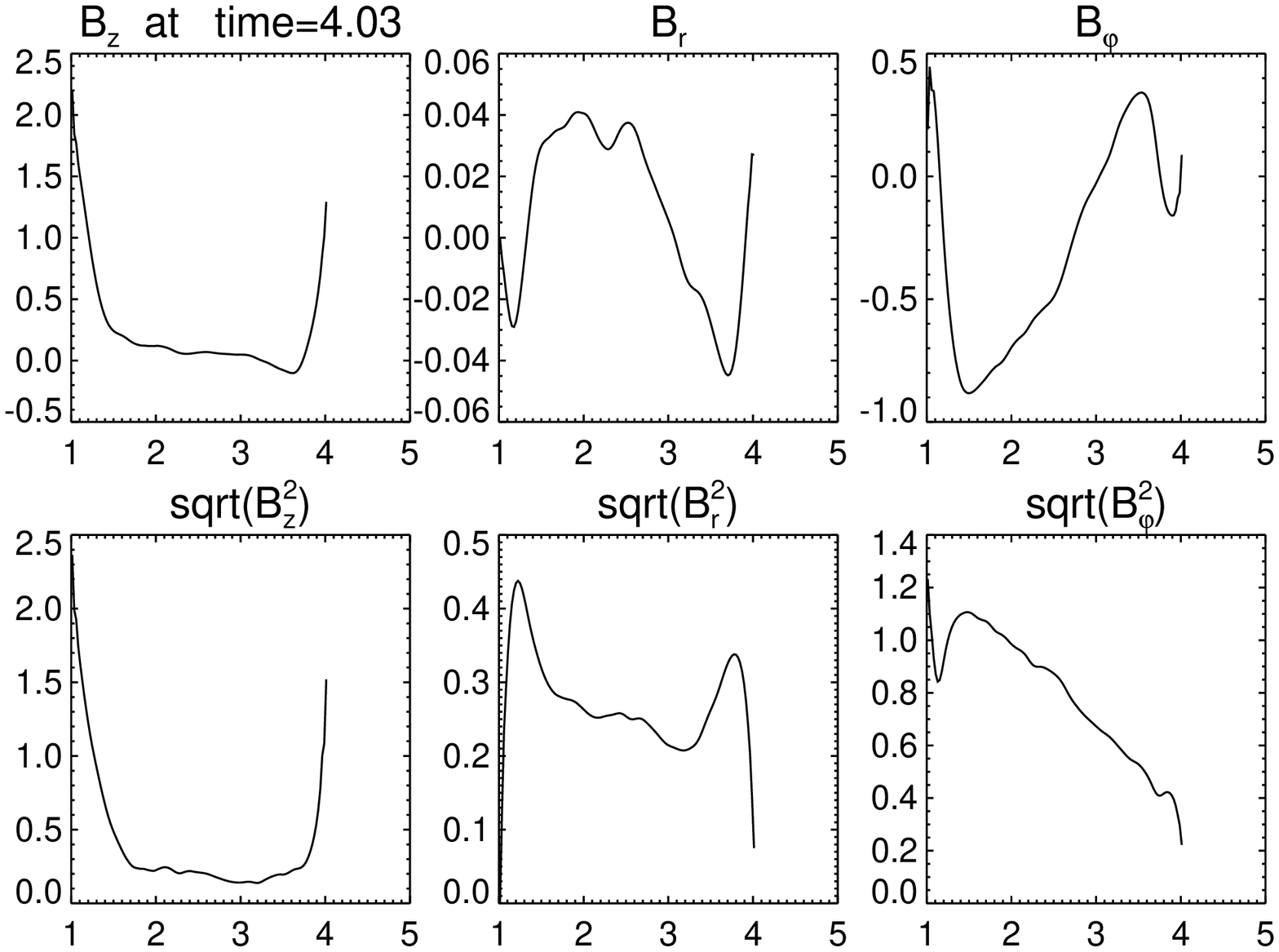}}
  \caption{Radial profiles of $\varphi,z$~averaged velocity components
            in [$\kms\,$] (first row), rms-velocity components in
            [$\kms\,$] (second row), $\varphi,z$-averaged magnetic
            field components in [$\muG\,$] (third row) and rms magnetic
            field components in [$\muG\,$] (fourth row) at $1.8\Gyr$ }
  \label{fig:mrirad}
\end{figure}
Applying a weak and uniform vertical magnetic field of $0.1\muG$, we find
an exponential growth of the unstable mode after a transient
relaxation phase of about $200\Myr$. The field starts to grow at the
inner radial boundary, where $\Omega$ has a maximum. As the field
grows further, it spreads out radially. At the same time, the Maxwell
stress causes an inflow in the inner part of the magnetised region. At
the outer magnetic region, we observe an outflow of mass, which is a
consequence of angular momentum conservation.

Because of our closed boundaries, the density increases strongly at
the inner boundary and moderately at the outer boundary. We see a weak
change of the rotation curve at the boundaries, compensating the
developing pressure gradients. This is probably an artefact of the
closed boundaries. Tests with larger radial domains have shown that
the change of the rotation curve is restricted to the boundary region.

Finally, the field growth saturates. At the moment it is unclear
whether the saturation is caused by the non-linear feedback of the
developed turbulence. The reduction of the pitch angle could in fact
be a hint that the system reaches the diffusive limit of the
instability, where we know from linear analysis that the unstable mode, 
in this regime, is mainly toroidal. On the other hand, we see a
large-scale meridional circulation in the disk. Moreover, our model
disk has a limited mass reservoir, which could be another reason for
the early saturation with rather low magnetic field strength. As long
as the toroidal and radial field strength are below the field strength
of the initial vertical field, the field parallel to the disk grows
with a time of $0.05\Gyr$; later we observe growth times of $0.8\Gyr$
until the saturation sets in. It should be noted that the growth-time
of a possible MRI-driven dynamo has not to be as fast as the growth
time of the instability.

In Fig.~\ref{fig:mrirad}, we plot radial profiles which have been
averaged along the vertical and azimuthal direction for
$-5\kpc < z < 0\kpc$ and $0 < \varphi < 2\upi$.
The values shown are the components of the velocity,
magnetic field and their corresponding rms values. A large-scale
meridional circulation develops with an amplitude of roughly 0.5 of
the rms value ($\simeq 6\kms$ for the vertical component and
$\simeq 3\kms$ for the radial component).

The mean magnetic field along the z and $\varphi$~direction are of
the same order as the rms~values of the fields. Only the mean radial
field is by an order of magnitude smaller than the rms~field. We find
$B_r^{\rm rms}/B_\varphi^{\rm rms}=0.3$ and for the mean-field
$B_r/B_\varphi=0.05 - 0.1$, corresponding to pitch angles of $17^\circ$
and $3^\circ - 6^\circ$ , respectively. In contrast, the pitch angle
in the polarisation map varies from $10^\circ$ to $20^\circ$ if the
boundaries are excluded ($1.3\kpc < r < 3.6\kpc$).

\begin{figure}[h]
\begin{center}
  \resizebox{4.2cm}{4.4cm}{\includegraphics{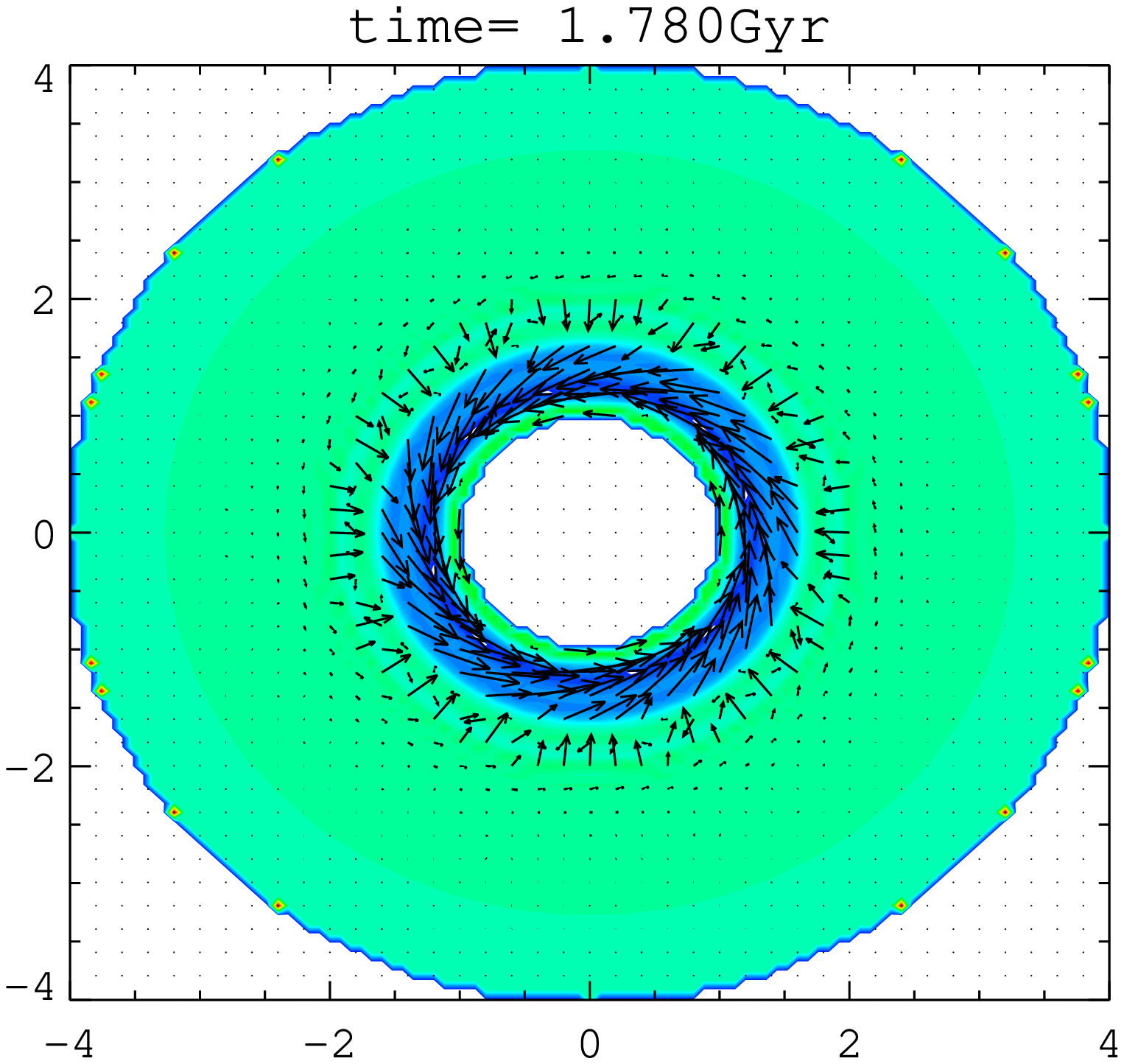}}
  \resizebox{4.2cm}{4.4cm}{\includegraphics{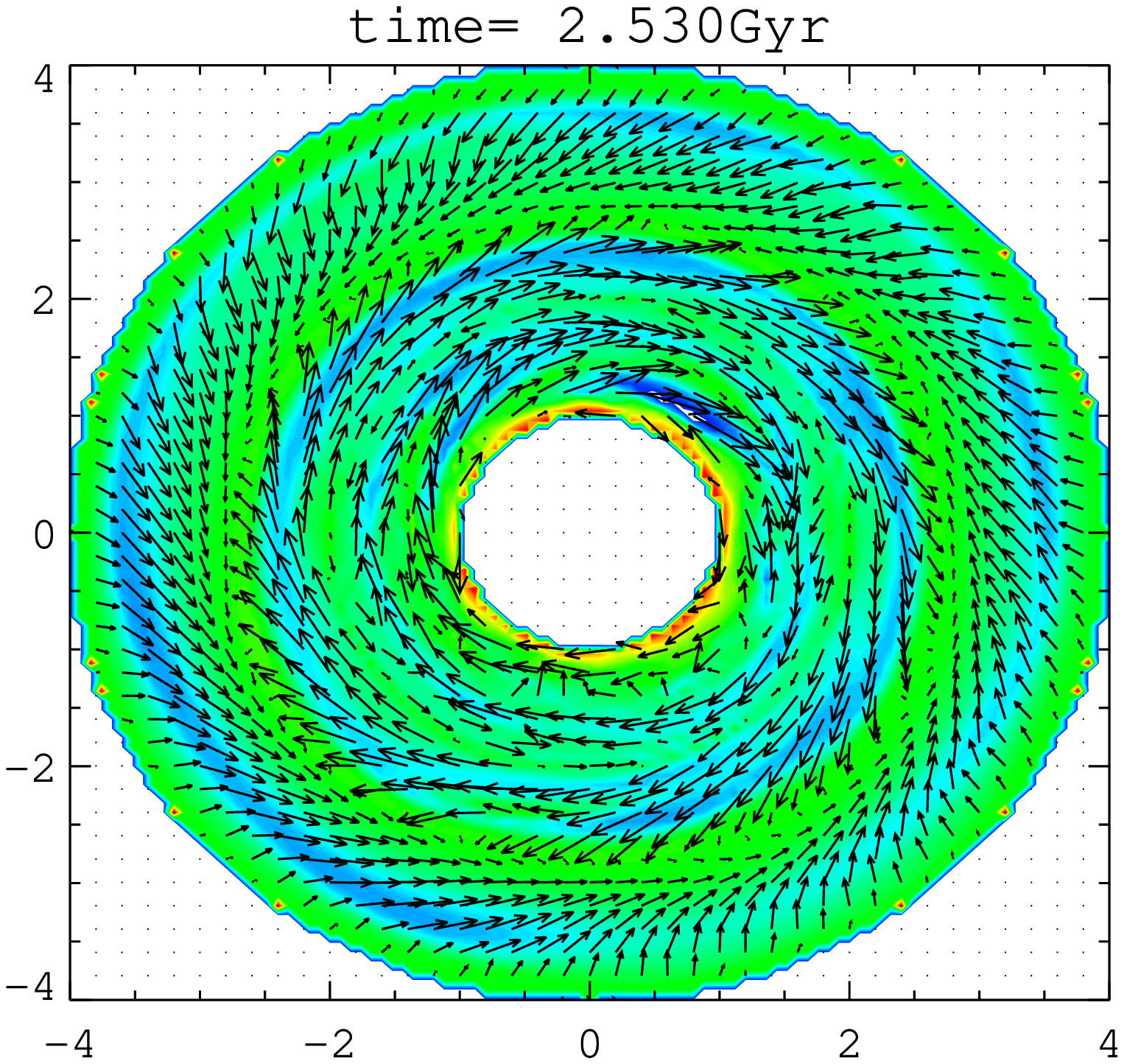}}
  \resizebox{4.2cm}{4.4cm}{\includegraphics{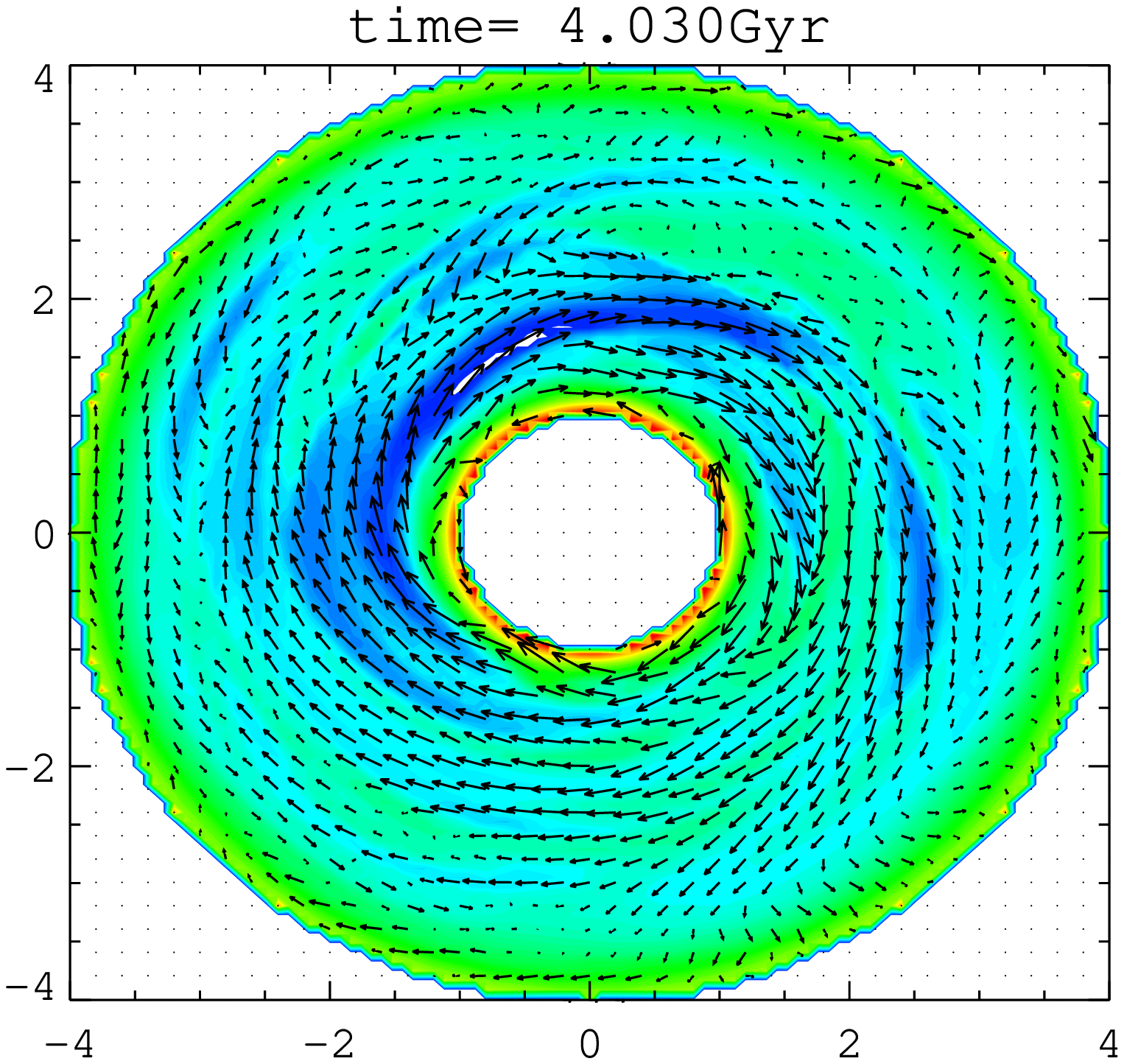}}

  \resizebox{4.2cm}{4.4cm}{\includegraphics{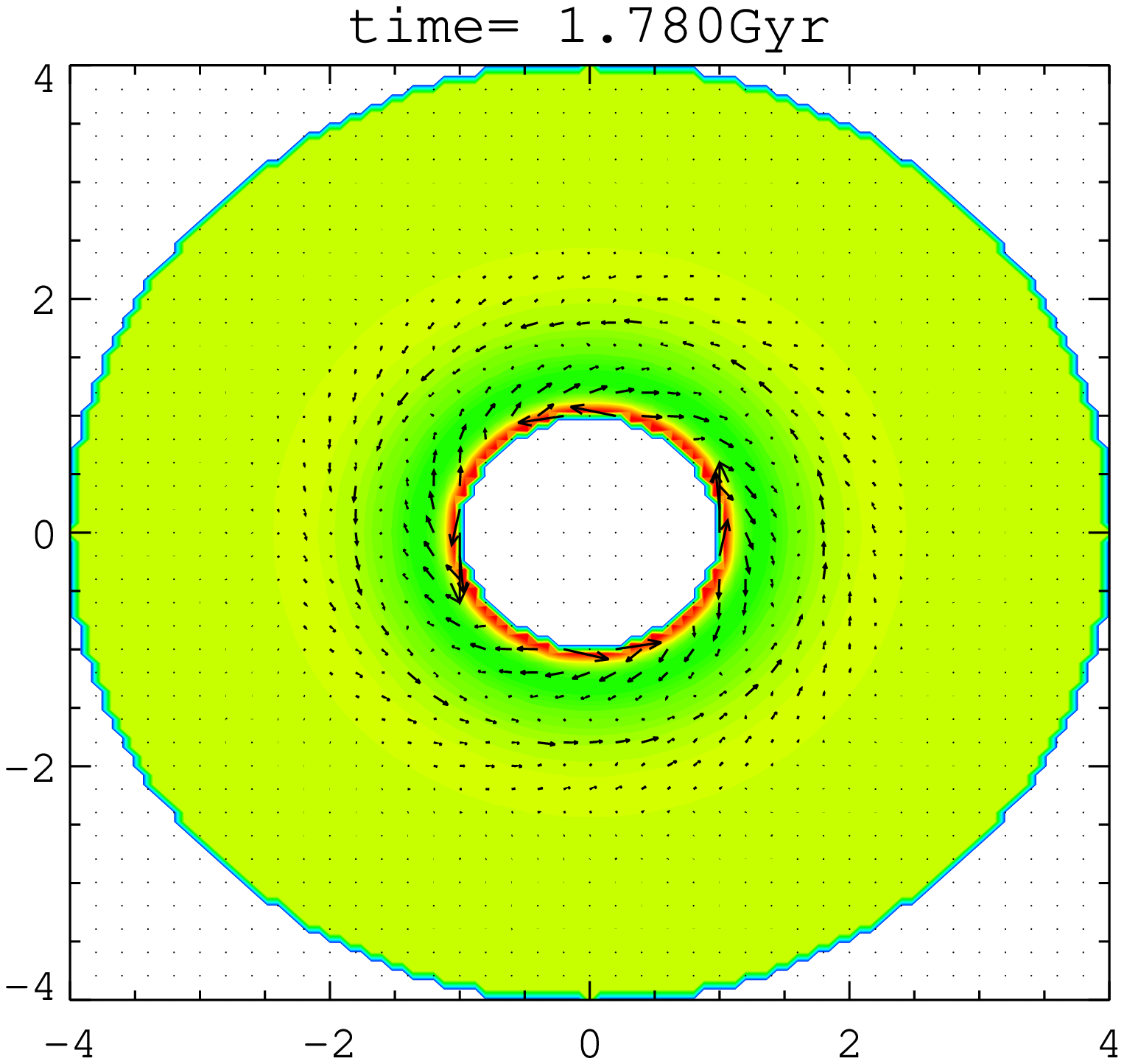}}
  \resizebox{4.2cm}{4.4cm}{\includegraphics{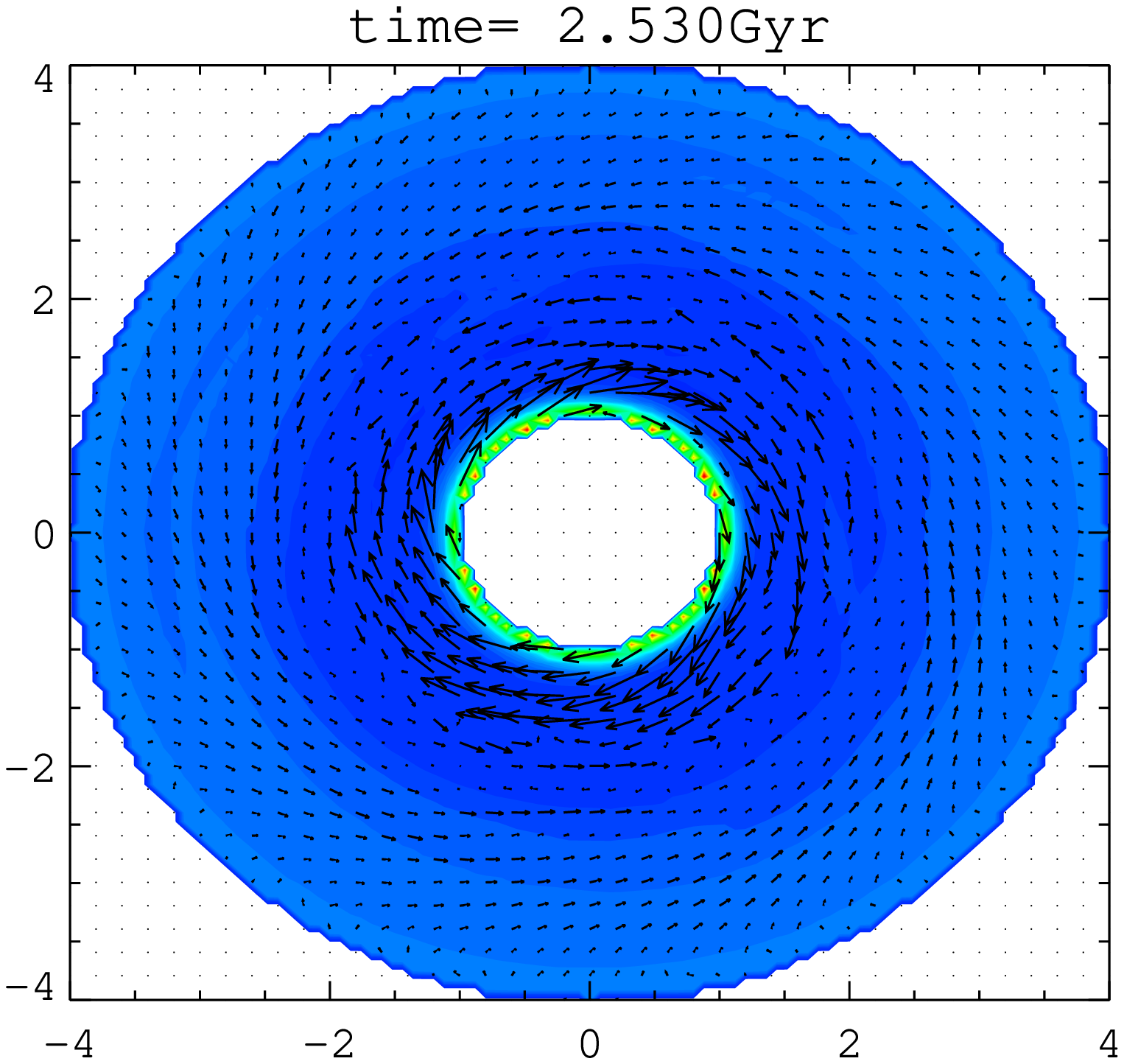}}
  \resizebox{4.2cm}{4.4cm}{\includegraphics{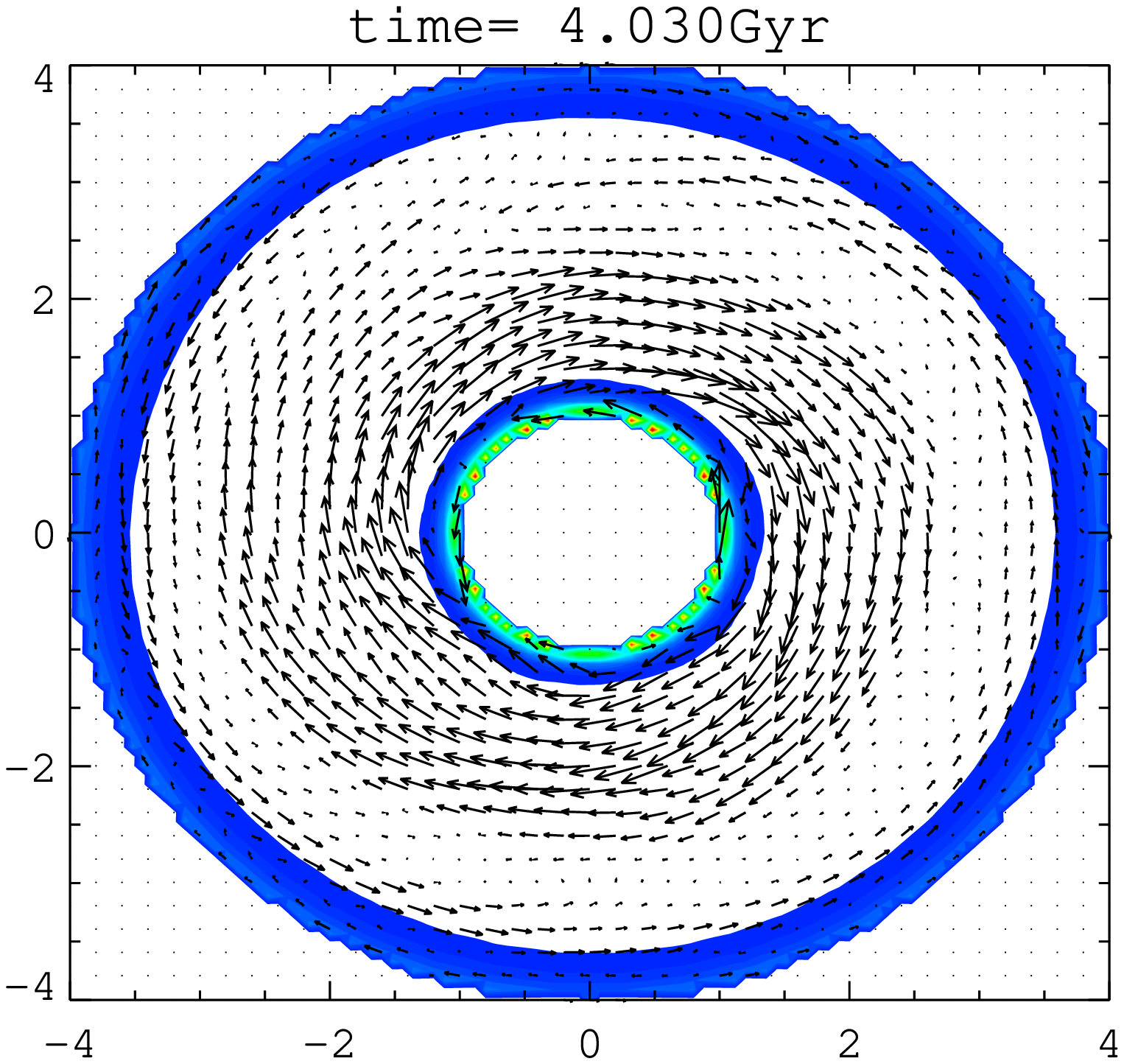}}

  \resizebox{4.2cm}{4.4cm}{\includegraphics{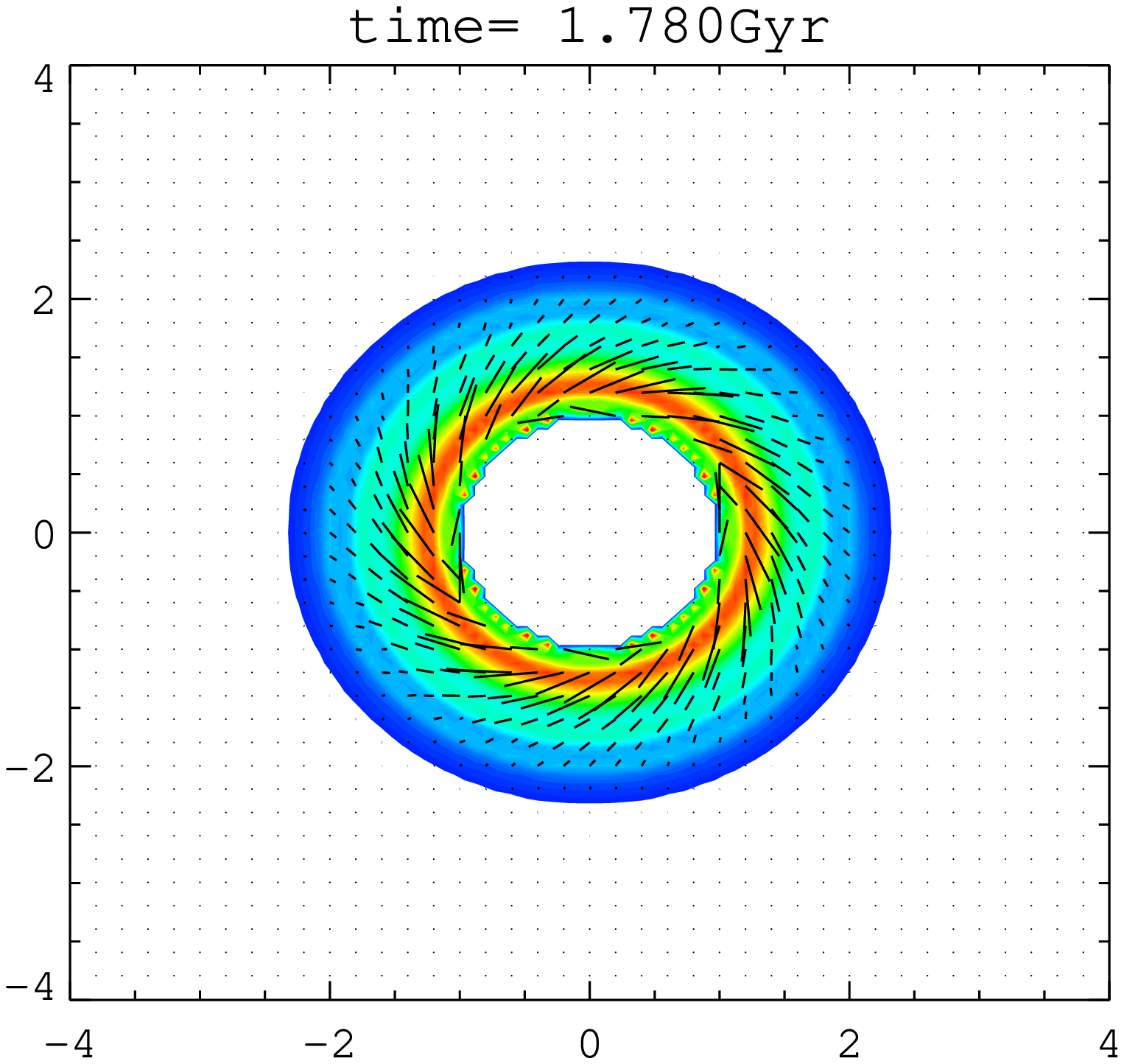}}
  \resizebox{4.2cm}{4.4cm}{\includegraphics{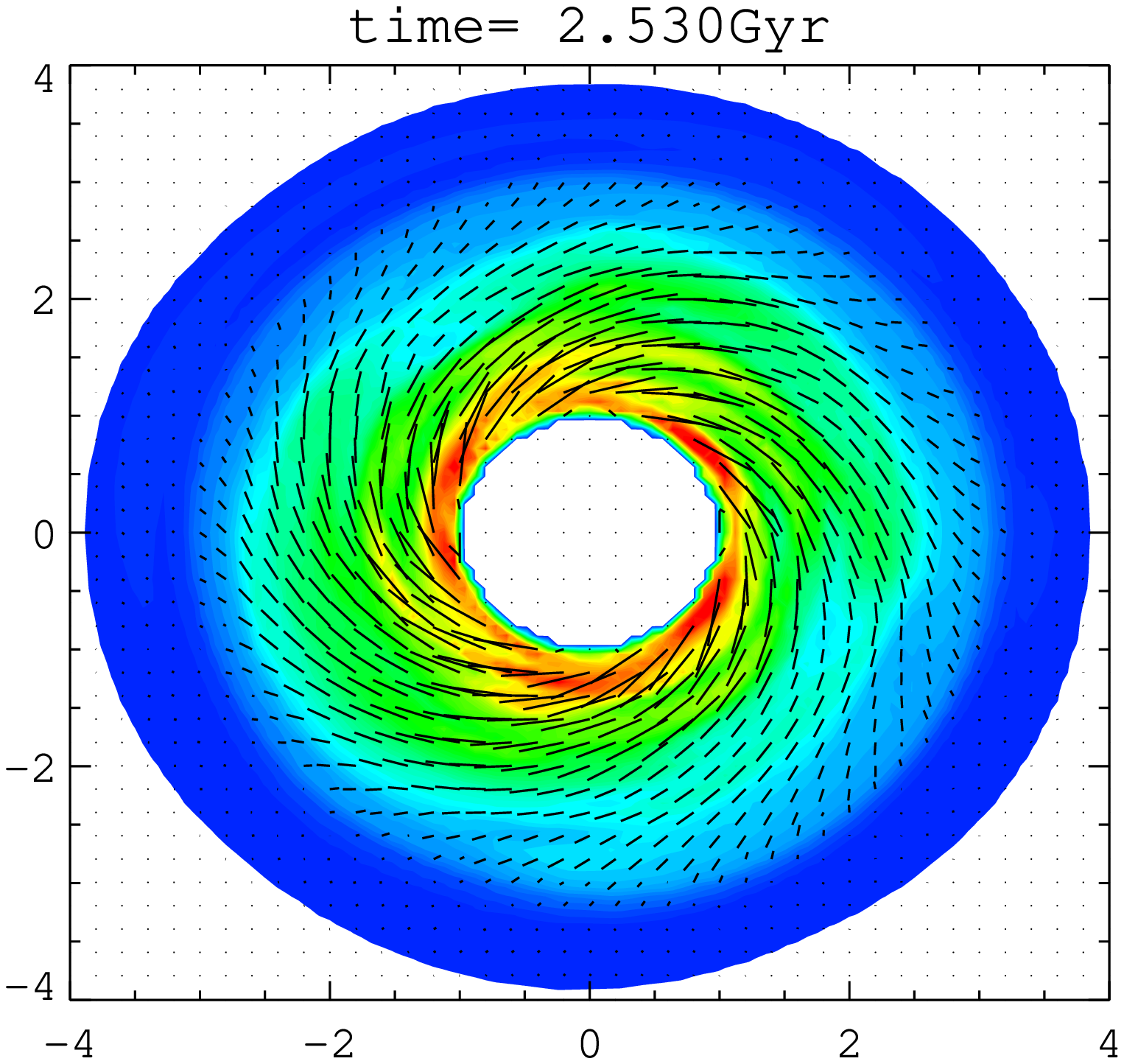}}
  \resizebox{4.2cm}{4.4cm}{\includegraphics{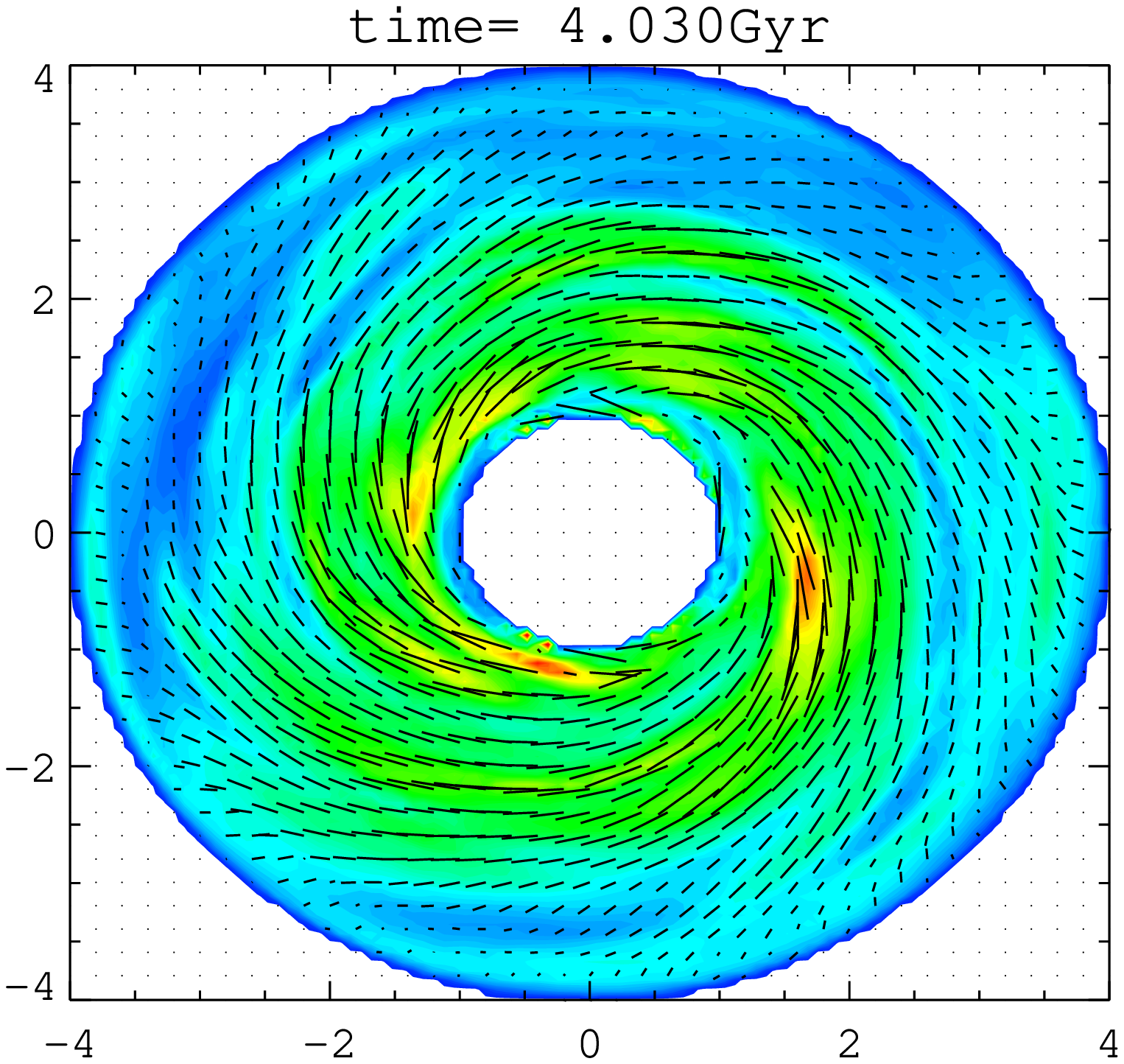}}
\end{center}
  \caption{Magnetic field slices at $z=-0.8\kpc$ (first row) and
  averaged field over one disk half (second row) with colour coded
  z~component. Magnetic field direction from polarised emission with
  colour coded total intensity (third row).}
  \label{fig:mridiskpol}
\end{figure}

Initially, at time $1.8\Gyr$, the pitch angle varies from $20^\circ$
to $40^\circ$ for $1.3\kpc < r < 2\kpc$. In the outer region the
field is still too weak, and we do not observe any polarised emission.
Fig.~\ref{fig:mridiskpol} shows a slice through the magnetic field at
$z=-0.8\kpc$, the vertically averaged field along the lower disk half
($-5\kpc< z < 0\kpc$) and the magnetic direction of polarised emission
(neglecting Faraday rotation). Varying the scale height height of the
relativistic electron distribution from $0.5\kpc$ to $2.0\kpc$ gave no
significant differences. Note the smooth polarisation map for an
incoherent magnetic field (i.e., with many reversals along the
vertical direction). The pitch angle is here much larger than the pitch angle
for the averaged field.

\section{Summary}\label{sec:summary}

New direct simulations of supernova-driven turbulence in a Cartesian
box have shown the existence of the turbulent dynamo in a galactic
environment. We find the resulting balance of the turbulent
diamagnetism with the mean vertical outflow to be crucial for the
operation of the dynamo. Using the turbulence of the direct
simulations for the determination of the transport coefficients, one
can show that the fast dynamo growth time of a few hundred million
years is due to the fast transport by the mean vertical fountain flow
and the downward pumping of the turbulent diamagnetism. This is in
contrast to the slow conventional $\alpha\Omega$~dynamo with growth
times of the order of $1\Gyr$, where only diffusion contributes to a
vertical field transport.

We have shown the possibility of an alternative saturation process
other than the simple quenching of the diagonal $\alpha$~term. In the
case of quenching via the transport mechanisms, the pitch angle of the
final saturated field will not deplete. But still, there are cases of
very strong differential rotation where it may be difficult to get
observed pitch angles of up to $40^\circ$ with actual SN-driven dynamo
models.

More detailed investigations of the ISM model must be performed. For
instance, cosmic rays could further improve the situation. Also the
effects of strong magnetic fields have to be considered, as magnetic
instabilities could contribute to further amplification processes.
The magneto-rotational instability would be a good candidate -- not
only as a generator for the seed field, but also for the generation of
magnetic fields with large pitch angles. It is not clear, however,
whether an MRI driven dynamo can generate enough vertical flux through
the disk in order to get solutions with large enough pitch angles,
which are known from the linear unstable modes. But at least under a
given vertical flux, the MRI leads to incoherent magnetic fields,
which appear in the polarised synchrotron emission (neglecting Faraday
rotation) as smooth spiral fields. The mean-field generated in such
models has a smaller pitch angle than the polarised emission vectors,
so a significant contribution from a small scale field enhances the
pitch angle in the polarisation map. Future RM~synthesis observations
would be very helpful for the understanding of magnetic fields in
external galaxies.


\begin{acknowledgments}
We thank N. Dziourkevitch and A. Bonanno for helpful discussions and
support concerning the MRI computations. We are also grateful to
U. Ziegler for supplying the NIRVANA code.
\end{acknowledgments}

%
%
%

\end{document}